# Field-induced hexagonal to square transition of the vortex lattice in overdoped $La_{1.8}Sr_{0.2}CuO_4$

R. Gilardi[1], J. Mesot[1], A.J. Drew[2], U. Divakar[2], S.L. Lee[2], N.H. Andersen[3], J. Kohlbrecher[4], N. Momono[5], M. Oda[5]

*(1) Laboratory for Neutron Scattering; ETH Zurich and PSI Villigen; CH-5232 Villigen PSI; Switzerland*
*(2) School of Physics and Astronomy; University of St. Andrews North Haugh; St Andrews KY16 9SS; United Kingdom*
*(3) Material Research Department; Riso National Laboratory; 4000 Roskilde; Denmark*
*(4) Spallation Neutron Source Division; PSI Villigen; CH-5232 Villigen; Switzerland*
*(5) Departement of Physics; Hokkaido University; Sapporo 060-0810; Japan*

ABSTRACT
We report on a small angle neutron scattering study of the vortex lattice in overdoped $La_{2-x}Sr_xCuO_4$ (x=0.2) up to high magnetic fields (9.5 Tesla) applied perpendicular to the $CuO_2$ planes. At low magnetic fields we observe a crossover from hexagonal to square coordination of the vortex lattice. This field-induced transition confirms the results obtained in slightly overdoped $La_{2-x}Sr_xCuO_4$ (x=0.17).

High-temperature superconductors (HTSC) renewed the interest in the study of vortex matter in Typ-II superconductors [1,2]. Surprisingly, almost no information exists about the microscopic observation of a vortex lattice (VL) in $La_{2-x}Sr_xCuO_4$ (LSCO), a compound belonging to the family of the first HTSC to be discovered. Using small angle neutron scattering (SANS), we have recently reported the first direct evidence of a well ordered VL in slightly overdoped LSCO (x=0.17) and the observation of a field-induced transition from a hexagonal to a square VL oriented along the Cu-O bonds [3,4]. The existence of an intrinsic square VL had never been observed in HTSC, and is indicative of the coupling of the VL to a source of in-plane anisotropy, such as those provided by the d-wave superconducting gap [5-7] or Fermi velocity anisotropies [8].

We report here SANS measurements on a more overdoped LSCO single crystal (x=0.20). The experiment was performed using the SANS-I diffractometer at the Paul Scherrer Institute, Switzerland. The high quality LSCO single crystal ($T_c$=31 K, m=1270 mg) was mounted in a cryostat in a magnetic field up to 9.5 Tesla (applied parallel to the incident beam) with the c-axis oriented along the field direction.

Fig.1 shows SANS diffraction patterns obtained at T=1.5 K after field-cooling in B=0.05 T, 0.4 T, 5 T and 9.5 T. A background measured above $T_c$ has been subtracted in order to remove the large signal arising from crystal defect scattering.

At very low fields (B=0.05 T) one observes a ring-like distribution of intensity, which is reminiscent of what we observed in slightly overdoped LSCO [3]. The diffraction pattern is interpreted as a superposition of diffraction from various domain orientations of hexagonal coordination. This interpretation is supported by the fact that the magnitude of the fundamental wavevector q of the VL at low magnetic fields is consistent to the expected value for a hexagonal VL (see Fig.2). As the field is increased from 0.05 to 0.4 Tesla the diffraction pattern completely changes, with the intensity concentrated in four spots forming a square lattice. Similarly to slightly overdoped LSCO the square VL is oriented in reciprocal space along the {1,1,0} directions (orthorhombic notation), which corresponds to the Cu-O bond directions [3]. By further increasing the magnetic field, the square VL remains oriented along the {1,1,0} directions (see Fig.1). At 9.5 Tesla some additional intensity is found along the {1,0,0} directions (between the four bright spots). Since at high magnetic fields the VL signal is very weak, this additional intensity most probably arises from a bad background subtraction. We can however not rule out that this intensity is intrinsic and indicate that at higher magnetic fields there is a reorientation of the square VL (rotation by 45º), as predicted by some theories [8,9], or a crossover to a more disordered (vortex glass?) phase. To solve this issue more experimental data at higher magnetic fields are needed.

Finally we would like to discuss quantitatively the hexagonal to square transition at low fields. The relationship between the magnetic field B and q depends on a structure dependent quantity $\sigma=(2\pi/q)^2 B/\Phi_0$, where $\sigma$ is equal to $\sqrt{3}/2$ for a hexagonal VL and 1 for a square one ($\Phi_0$ is the flux quantum) [3]. Fig.2 shows that the experimental values of $\sigma$ above B≈0.4 T are as expected for a square VL, while at low fields (B≤0.1T) they are consistent with a hexagonal VL. At intermediate magnetic fields (0.1T≤B≤0.4T) we probably have a superposition of domains with hexagonal and square coordination. Alternatively we can look at the ratio between the intensity along the {1,0,0} and the {1,1,0} directions. At low fields we have a ring-like intensity and therefore $Int_{\{1,0,0\}}/Int_{\{1,1,0\}}\approx 1$ while at higher fields the intensity is concentrated along the {1,1,0} direction and consequently $Int_{\{1,0,0\}}/Int_{\{1,1,0\}}$ is sharply reduced (see Fig.2).

In conclusions, our small angle neutron scattering study of the vortex matter in overdoped LSCO reveals a field-induced hexagonal to square transition of the VL. The square VL is oriented along the Cu-O bond direction up to the highest magnetic field applied (9.5 Tesla). These results are consistent with data obtained on a slightly overdoped sample and are therefore intrinsic features of the overdoped regime of LSCO.

*This work was performed at the spallation neutron source SINQ, Paul Scherrer Institut, Villigen, Switzerland, and was supported by the Swiss National Science Foundation, the Engineering and Physical Science Research Council of the U.K., the Ministry of Education and Science of Japan, and the Danish Technical and Natural Science Research Councils.*

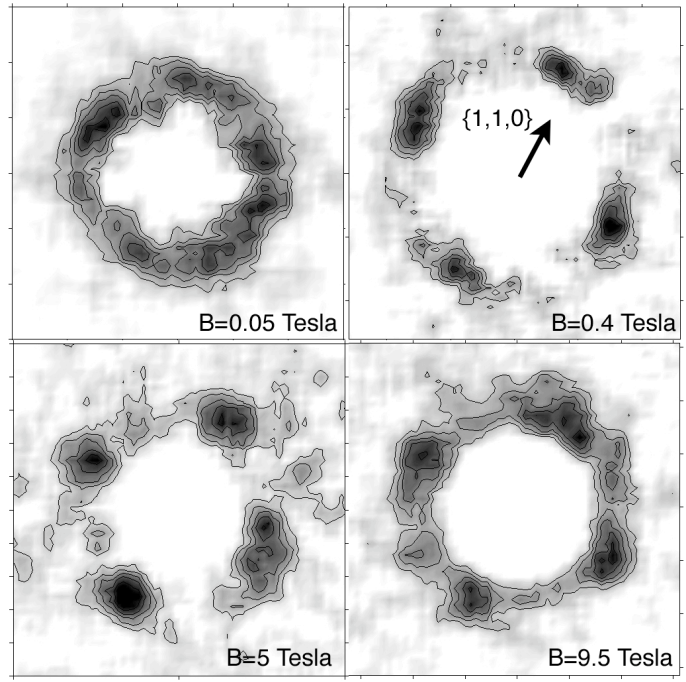

Fig.1: SANS diffraction patterns from the vortex lattice in overdoped LSCO (x=0.20), measured at T=1.5 K at an applied magnetic field of B=0.05 T, 0.4 T, 5T and 9.5 T.

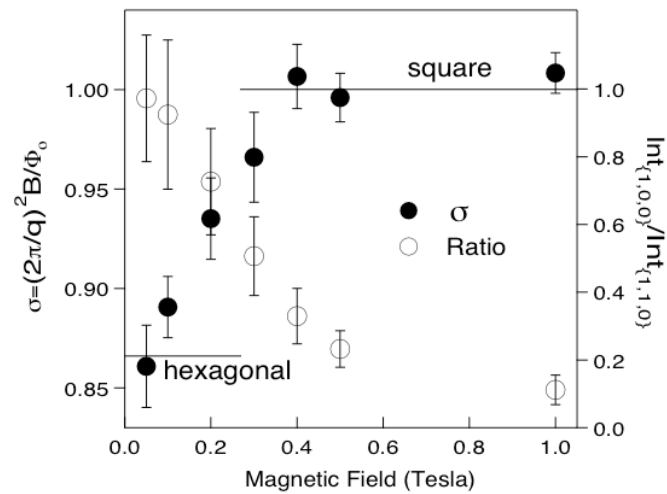

Fig.2: Quantitative analysis of the field-induced hexagonal to square transition. The full circles indicate the experimental value of $\sigma$ as a function of field, while the empty circles represent the intensity ratio between the {1,0,0} and {1,1,0} directions.